\documentclass[prl,twocolumn,showpacs,preprintnumbers,amsmath,amssymb,tightenlines,superscriptaddress]{revtex4}

\usepackage{graphicx}
\usepackage{dcolumn}
\usepackage{bm}
\usepackage{epsfig}
\usepackage{stmaryrd}
\usepackage{tabularx}
\usepackage{bbold}

\newcommand{\ssection}[1]{{\noi  \it #1:}}
 
\newcommand{\sub}[2]{{#1}_{\mbox{\!\! \scriptsize #2}}}

\def\noi{\noindent}
\def\beq{\begin{equation}}
\def\eeq{\end{equation}}

\def\CR{\nonumber\\[0.15cm]}

 \newcommand{\bra}[1]{\langle\,{#1}\, |}
 \newcommand{\ket}[1]{|\,{#1}\,\rangle}
\newcommand{\braket}[2]{\mbox{$\langle\,{#1}\, | \,{#2}\,\rangle$}}

\newcommand{\rref}[1]{Ref.~\cite{#1}}
\newcommand{\fref}[1]{Fig.~\ref{#1}}
\newcommand{\frefp}[2]{Fig.~\ref{#1}~(#2)}

\newcommand{\eref}[1]{Eq.~(\ref{#1})}

\newcommand{\cref}[1]{chapter~\ref{#1}}

\newcommand{\Cref}[1]{Chapter~\ref{#1}}
\newcommand{\tref}[1]{table~\ref{#1}}

\newcommand{\bref}[1]{(\ref{#1})}

\usepackage{ulem}  
\usepackage{color}

\begin{document}

\title{A measure for adiabatic contributions to quantum transitions}
\author{R.~Pant}
\affiliation{Department of Physics, Indian Institute of Science Education and Research, Bhopal, Madhya Pradesh 462 066, India}
\email{ritesh17@iiserb.ac.in}
\author{P.~K.~Verma}
\affiliation{Department of Chemistry, Indian Institute of Science Education and Research, Bhopal, Madhya Pradesh 462 066, India}
\author{C.~Rangi}
\affiliation{Department of Physics, Indian Institute of Science Education and Research, Bhopal, Madhya Pradesh 462 066, India}
\author{E.~Mondal}
\affiliation{Department of Chemistry, Indian Institute of Science Education and Research, Bhopal, Madhya Pradesh 462 066, India}
\author{M.~Bhati}
\affiliation{Department of Chemistry, Indian Institute of Science Education and Research, Bhopal, Madhya Pradesh 462 066, India}
\author{V.~Srinivasan}
\affiliation{Department of Chemistry, Indian Institute of Science Education and Research, Bhopal, Madhya Pradesh 462 066, India}
\email{vardha.ac.in}
\author{S.~W\"uster}
\affiliation{Department of Physics, Indian Institute of Science Education and Research, Bhopal, Madhya Pradesh 462 066, India}
\email{ sebastian@iiserb.ac.in}
\begin{abstract}
We construct a measure for the adiabatic contribution to quantum transitions in an arbitrary basis, tackling the generic complex case where dynamics is only partially adiabatic, simultaneously populates several eigenstates  and transitions between non-eigenstates are of key interest. 
Our measure is designed to distinguish transitions between basis states that occur due to the adiabatic change of the underlying populated eigenstates from transitions that occur due to beating between several such eigenstates. We demonstrate that the measure can be applied to material or molecular simulations using time-dependent density functional theory, allowing to quantify the relative importance of adiabaticity and thus nuclear motion, for example, in charge or energy transfer.
\end{abstract}

\maketitle

\ssection{Introduction}
%
Adiabatic changes of selected quantum states via slow manipulation of the Hamiltonian are essential for quantum state control and understanding complex quantum dynamics. They are of key utility in adiabatic quantum computation \cite{albash2018adiabatic, gosset2015universal, sarandy2005adiabatic,Phys_adiabgate_PhysRevA}, quantum optimisation \cite{steffen2003experimental}, chemical reactions \cite{ButlerARPC1998,DavidScience1998,FernandoJPCB2021,KazuoBCSJ2021,WeibinSA2022}, nuclear motion in photo-chemistry \cite{may2011charge} and quantum state preparation schemes \cite{chen2012long, chen2016robust,eckert2007efficient} such as stimulated Raman adiabatic passage (STIRAP) \cite{vitanov_stirap_RevModPhys}.

Despite this importance for quantum technologies and understanding dynamical processes, no generic measure for adiabaticity exists, while they were proposed for entanglement \cite{Vedral_quant_entanglem_PhysRevLett}, coherence \cite{Baumgratz_Coherence_PhysRevLett} and non-Markovianity \cite{Breuer_NonMarkovMeasure_PRL}, for example. As long as one starts in one fixed initial state, transforming it into a given target state, the net change of eigenstate populations can be used. This can constrain when evolution due to a certain Hamiltonian should be fully adiabatic, although this remains non-trivial to predict \cite{Comparat_general_adiab_PhysRevA,Jiangfeng_exp_adiab_PhysRevLett,Marzlin_inconsist_PhysRevLett}.

In more complex scenarios, the initial state will be a superposition of eigenstates and dynamics will give rise to frequent non-adiabatic effects. Then
 it becomes highly nontrivial to quantify to what extent transitions between non-eigenstates are due to adiabatic evolution or would also have occurred with a constant Hamiltonian owing to interference between eigenstates. Exemplary scenarios include the quantum transport of an electronic excitation in a molecular aggregate \cite{2013photonics, dijkstra2019efficient, caruso2009highly} through molecular motion \cite{pant2020excitation,Asadian_2010,semiao2010vibration,behzadi2017effects,o2014non,mulken2011directed, rehhagen2022effect} or vibrationally assisted intersystem crossing and charge transfer \cite{EvansJPCL2018,SerdiukJPCB2021,DasJPCA2022,Stier_Nonadiab_electransfer_JPCB}.

\begin{figure}[htb]
\includegraphics[width=0.99\columnwidth]{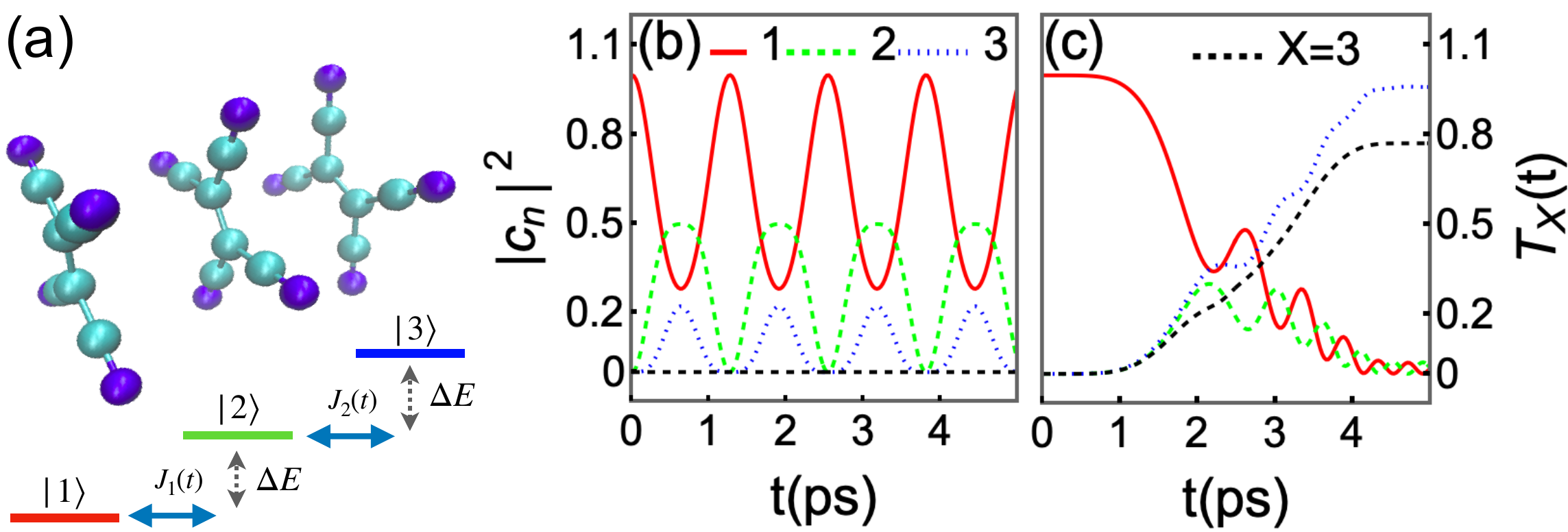}
\caption{(a) Molecular trimer aggregate with on-site energy shifts and three effective diabatic states $\ket{N}$, representing e.g.~localized energy or charge on monomer $N$. 
(b) Quantum transitions due to beating. We show populations $p_n=|\braket{n}{\Psi(t)}|^2$ for $n=1,2,3$ as per legend, for constant transition matrix elements $J_k(t)=J_0$, 
using $\Delta E= 3$ and $J_0=2$. The black dashed line in (b,c) is the adiabatic transport measure $T_3(t)$. (c) Quantum transitions due to adiabatic changes in the Hamiltonian, while varying $J_k(t)$ as discussed in the text. $\Delta E$ and legend are as in (b). 
\label{sketch_problem_statement}}
\end{figure}
We address this challenge by constructing a general measure to discriminate whether transitions in an arbitrary basis are caused by time-evolving eigenstates or rather by beating between super-imposed eigenstates. We derive the measure from the time-dependent Schr{\"o}dinger equation (TDSE), illustrate its functionality on diverse dynamical examples, and then demonstrate its full potential by an application to charge transport in time-dependent density functional theory (tdDFT).

\ssection{Problem statement}
%
Consider a time-dependent Hamiltonian $\hat{H}(t)$ with a discrete spectrum and let $\{\ket{n}\}$ be an arbitrary, time-independent, ortho-normal basis of the Hilbert space, the diabatic basis. Meanwhile $\ket{\varphi_k(t)}$ is a solution of the instantaneous eigenproblem
$
\hat{H}(t)\ket{\varphi_k(t)}=U_k(t)\ket{\varphi_k(t)},
$
with energy $U_k(t)$. The $\ket{\varphi_k(t)}$ form the adiabatic basis. The time-evolving state $\ket{\Psi(t)}$ can be expressed in either basis as
$\ket{\Psi(t)}=\sum_n c_n(t) \ket{n}$ or $\ket{\Psi(t)}=\sum_k \tilde{c}_k (t) \ket{\varphi_k(t)}$, with expansion coefficients related by $\tilde{c}_k (t)  =  \sum_n d^{(k)*}_n(t) c_n(t)$, where $d^{(k)}_n(t)=\braket{n}{\varphi_k(t)}$ for the amplitude of $\ket{n}$ contained in eigenstate $\ket{\varphi_k(t)}$. For idealized cases with near constant adiabatic populations $\tilde{p}_k=|\tilde{c}_k(t)|^2$, 
joint inspection of diabatic populations $p_k=|c_k|^2$ can suffice to assess adiabaticity of a transition. However, much more generally, multiple $\tilde{p}_k$ will be non-zero and many will vary, making it challenging to assess whether or not a given diabatic transition $p_k\rightarrow p_{k'}$ has been aided by adiabatic state following, was hindered by it, or would have occurred for constant Hamiltonian.

Suppose we seek the physical root cause of population changes in the basis $\{\ket{n}\}$ instead of $\{\ket{\varphi_k(t)}\}$, for example, because $\{\ket{n}\}$ is spatially localized to describe transport. These populations are
\begin{align}
p_n(t)&=|c_n(t)|^2=\sum_{k,k'} \tilde{c}^*_{k'}(t)\: \tilde{c}_k(t) \:d^{(k)}_n(t)\: d^{(k')*}_n(t).
\label{pop}
\end{align}
Now there are three distinct contributions to changes in $p_n(t)$: (i) For a time independent Hamiltonian, we would have $d^{(k)}_n=const$ and $\tilde{c}_k(t) = \tilde{c}_k(0) \exp{[-i U_k t/\hbar]}$, nonetheless populations $p_n$ may vary in time due to \emph{beating} from interference terms containing $e^{-i \Delta E_{kk'} t/\hbar }$ with $\Delta E_{kk'}=E_{k'}-E_k$. (ii) For a time dependent Hamiltonian,
populations can change also due to variations of the $\ket{\varphi_k(t)}$ affecting the $d^{(k)}_n(t)$, clearly an \emph{adiabatic} contribution, or (iii)  \emph{non-adiabatic} changes in $|\tilde{c}_k(t)|$.

All the cases are illustrated in \fref{sketch_problem_statement} for a three-level system that could be the trimer aggregate shown, with Hamiltonian 
$\hat{H}(t)=\sum_n E_n \ket{n}\bra{n} + \sum_{nm} J_{nm}(t)\ket{n}\bra{m}$.
When only nearest neighbor couplings exist and are constant, $J_{n(n+1)}(t)=J_0$, with system initialized in $\ket{1}$,
it eventually reaches state $\ket{3}$ due to beating, see panel (b). When couplings vary as 
$J_{12}(t) = \bar{J}_{12} \sin^2{(\pi t/2 \sub{t}{max})}$ and $J_{23}(t) = \bar{J}_{23} \cos^2{(\pi t/2 \sub{t}{max})}$ with $\bar{J}_{nm} = 8$, the system reaches $\ket{3}$ 
\emph{only} due to the temporal change of the Hamiltonian (akin to STIRAP), and never would for $J_{nm}(t)\equiv J_{nm}(0)$. In generic complex quantum dynamics both phenomena co-exist,
yet one might wish to quantify to what extent adiabaticity contributes, for example, to optimize a target chemical reaction \cite{DavidScience1998}
or electron transfer from a molecular dye donor to a semi-conductor acceptor \cite{Stier_Nonadiab_electransfer_JPCB}.

\ssection{Measure for the adiabaticity of transitions}
%
To tackle this problem, we split the population change $\dot{p}_n(t)$ into
\begin{align}
&\dot{p}_n(t)=\sum_{k,k'}\big[ \left(\dot{\tilde{c}}^*_{k'}(t)\: \tilde{c}_k(t) +  \tilde{c}^*_{k'}(t)\: \dot{\tilde{c}}_k(t)\right) \:d^{(k)}_n(t)\: d^{(k')*}_n(t)\CR
&\!\!\!\!\!\!+ \tilde{c}^*_{k'}(t)\: \tilde{c}_k(t) \: \left( \dot{d}^{(k)}_n(t)\: d^{(k')*}_n(t) + d^{(k)}_n(t)\: \dot{d}^{(k')*}_n(t) \right) \big].
\label{pop_dot}
\end{align}
The last line clearly already contains contributions to $\dot{p}_n(t)$ only from temporal changes of $\ket{\varphi_k(t)}$ and thus will be related to the adiabatic state following.
However, in scenarios where $\tilde{c}_k(t)$ are not constant, this can be modified by contributions from the first line, see SI, that were not considered in \rref{Stier_Nonadiab_electransfer_JPCB}.
For a time-independent Hamiltonian we could use  $\tilde{c}_k(t) = \tilde{c}_k(0) \exp{[-i U_k t/\hbar]}$, to reach
\begin{align}
&\dot{\tilde{c}}^*_{k'}(t)\: \tilde{c}_k(t) +  \tilde{c}^*_{k'}(t)\: \dot{\tilde{c}}_k(t)
= i \frac{\Delta E_{kk'} }{\hbar}\tilde{c}^*_{k'}(0)\: \tilde{c}_k(0) e^{i \Delta E_{kk'} t/\hbar },
\label{Vdd}
\end{align}
quantifying the temporal changes of $p_n(t)$ due to beating between different eigenstates, where $\Delta E_{kk'}=U_k' - U_k$. In that case, only
the phase of \bref{Vdd} is time-dependent, not the modulus.

We now exploit this for the isolation of contributions from adiabatic population changes, and write the coefficient $\tilde{c}_k(t)$ in polar representation $\tilde{c}_k(t)=\tilde{a}_k(t) e^{i \tilde{b}_k(t)}$, with $\tilde{a}_k,\tilde{b}_k\in \mathbb{R}$, $\tilde{a}_k>0$, such that $\dot{\tilde{c}}_k(t)=\dot{\tilde{a}}_ke^{i \tilde{b}_k(t)} + \tilde{a}_k(t) [i\dot{\tilde{b}}_k]e^{i \tilde{b}_k(t)}$. 
After insertion into \bref{pop_dot}, we remove the phase evolution $\dot{\tilde{b}}_k$ and define the remainder
\begin{align}
\label{pop_dot_no_phase}
f_n(t)=&\sum_{k,k'}\bigg[ \bigg(\dot{\tilde{a}}_{k'}e^{-i \tilde{b}_{k'}} \: \tilde{c}_k +  \tilde{c}^*_{k'}\:\dot{\tilde{a}}_ke^{i \tilde{b}_k}  \bigg) \:d^{(k)}_n\: d^{(k')*}_n\CR
&+ \tilde{c}^*_{k'}\: \tilde{c}_k \: \left( \dot{d}^{(k)}_n\: d^{(k')*}_n + d^{(k)}_n\: \dot{d}^{(k')*}_n \right) \bigg],
\end{align}
where the time argument $t$ of all functions on the RHS is suppressed.
The real variable $f_n(t)$ measures the rate of change of the population in state $n$ due to temporal changes in the eigen-spectrum of the Hamiltonian only,
by construction not containing any contribution from beating between multiple occupied eigenstates.

The best final assembly of a measure for the adiabatic contribution to transitions into a target state $\ket{X}$ will depend on the physical scenario of interest.
Here we shall use the integral of \bref{pop_dot_no_phase}, $T_X(t)=  \int_0^t  dt' f_X (t')$, for demonstrations. Another proposal is described in the SI \cite{sup:info}.
We verified in \fref{sketch_problem_statement} that $T_X(t)$ can successfully discriminate adiabatic transitions from beating. It remains zero for beating in (b), and approaches 
the transferred target population $T_X(t)\rightarrow p_X$ for a strongly adiabatic transition, reaching $T_X(t)=p_X$ if the transition was exclusively adiabatic.
We will now benchmark the measure on a more complex set of examples, representing the problem we intend to address. 

\begin{figure}[htb]
\includegraphics[width=0.99\columnwidth]{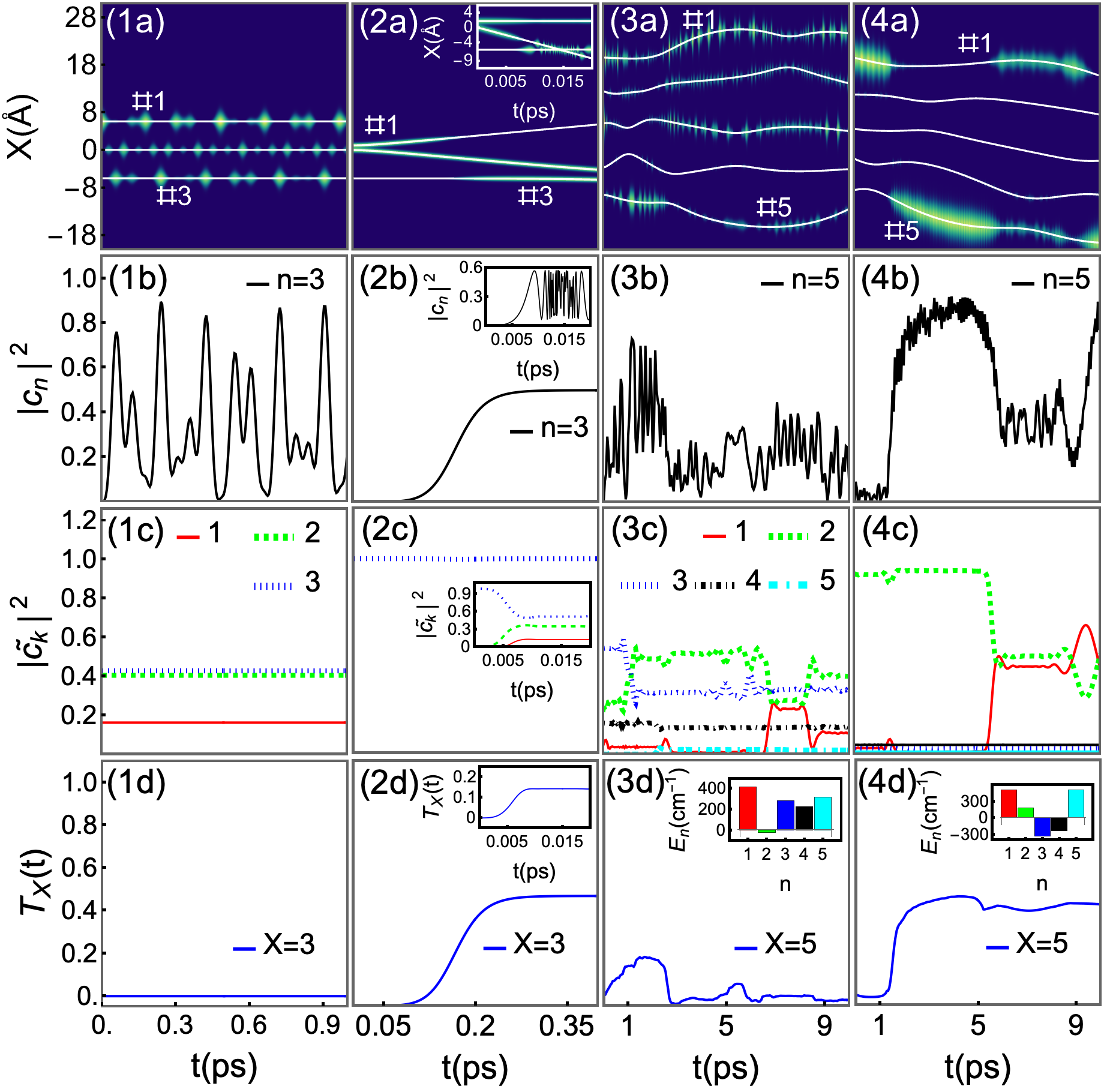}
\caption{Exciton transport in molecular aggregates. The first row (1a - 4a) shows the trajectories $X_k(t)$ of individual molecules (white lines) with the excitation probabilities of each molecule $p_n=|c_n|^2$ (diabatic populations) indicated by the width of color shading.  (1b-4b) isolated population $p_X$ on the target site only. The third row (1c-1d) shows adiabatic populations $\tilde{p}_n=|\tilde{c}_n|^2$ and the fourth row (1d-4d) the proposed adiabaticity measure for the target site $T_X$ (solid line). The columns differ by initial state and parameters as discussed in the text. The insets in column 2 are for a similar scenario with faster motion and thus less adiabatic dynamics.
\label{fig_trimer_cradles}}
\end{figure}

\ssection{Adiabatic excitation transport in molecular aggregates}
%
For this, we consider excitation energy transport in molecular aggregates in which monomers are mobile. The aggregate Hamiltonian is 
\begin{align}
\label{molagg_hamil}
\sub{\hat{H}}{agg}& = \sum_{n=1}^{N}E_{n} \ket{n}\bra{n} + \sum\limits_{n\neq m}\frac{ \mu^2}{|X_{mn}|^3} \ket{n}\bra{m},
\end{align}
where now $\ket{n}$ implies a single exciton state where all monomers are in their electronic ground state, but the $m^{th}$ one excited. This excitation can then migrate over the aggregate through transition dipole-dipole interactions of strength $\mu^2$ and since these depend on the separation $X_{nm}(t)=X_m(t)-X_n(t)$ of monomers $n$ and $m$, the Hamiltonian becomes time dependent. We solve the TDSE to simulate exciton dynamics, while monomers move classically through Newton's equation and interact via a Morse potential \cite{sup:info}. It was shown in 
\rref{pant2020excitation}, that motion can help overcome localization of the excitation due to disorder $E_n$, which arises through interactions with an environment \cite{2013photonics, kunsel2021scaling, bondarenko2020exciton}. For simulations shown in \fref{fig_trimer_cradles}, we have taken $\mu = 1.12$ a.u., and $M= 902330$ a.u., roughly matching carbonyl-bridged triaryl-amine (CBT) dyes \cite{saikin2017long}, with disorder realisations $E_n$ shown in the inset of row (d) where applicable.

The first column (1a-1d) shows an immobile case with $M\rightarrow \infty$ without disorder. The initial state is $\ket{1}$, which is not an eigenstate, and population quickly reaches $\ket{3}$ in the resultant beating, contribution (i) below \eref{pop}. Per construction, the adiabaticity measure $T_3(t)$ remains zero, as can be seen in (1d). 
 In the second column (2a-2d), monomers are mobile and the excitation initially shared, 
 $
\ket{\psi(0)} =  (\ket{n=1} + \ket{n=2})/\sqrt{2}.
$
The excitation reaches the output site solely due to adiabatic quantum state following, contribution (ii), as in \cite{wuster2010newton, mobius2011adiabatic}. Hence all population remains constantly in the initially occupied eigenstate, as shown in (2c). Once the excitation has reached the output site with probability $p=1/2$ at about $t=0.3$ ps, also the measure $T_3=1/2$, indicating that transport has been entirely adiabatic. For the same scenario, but made less adiabatic by faster motion of the second molecule, we see in the (2c) inset that the adiabatic population has dropped to $0.5$ by the time $t=0.01$ ps. The adiabatic measure in the inset of (2d) accordingly decreased, compared to the ideal adiabatic transport in the main panel. 

The remaining two columns show key cases for which the measure was developed: complex quantum dynamics in thermally agitated aggregates with disorder. For the latter we distribute $E_n$ according to $p_E(E_n) = \frac{1}{\sqrt{2\pi \sigma_E}} e^{-E_n^2/(2\sigma_E^2)}$. Initial positions and velocities are thermally distributed for temperature  $T=300$ K and the excitation initialised in $\ket{1}$. Columns 3 and 4 differ by disorder realization and inter-monomer potential, affecting the character of motion, see \cite{sup:info}. In both cases, site $\ket{5}$ is reached with a significant probability, while adiabatic populations are constant only for finite intervals between significant changes, thus dynamics is merely partially adiabatic. 

To assess manually how strongly adiabatic state following has contributed to the excitation arriving on site 5 now requires careful inspection of populations in both cases. One then sees in column 4, that the arrival at $t\approx1.5$ ps with population conserved in a single eigenstate was clearly adiabatic, while the departure at $t=5$ ps involved non-adiabatic splitting onto two eigenstates, contribution (iii), and subsequent beating between these. Accordingly, $T_5$ changes at the arrival time of population $|c_X|^2$, between $1-2$ ps, but not at the departure time, around $5$ ps. The main contribution to population on site 5 in column 3 after 3 ps is from beating, since the initial adiabatic contribution has been adiabatically removed at $t=3$ ps, according to $T_5$. Now, the measure removes the need for such detailed scrutiny of the dynamics, can be evaluated automatically and thus allows ensemble averages.

\begin{figure}[htb]
\includegraphics[width=0.99\columnwidth]{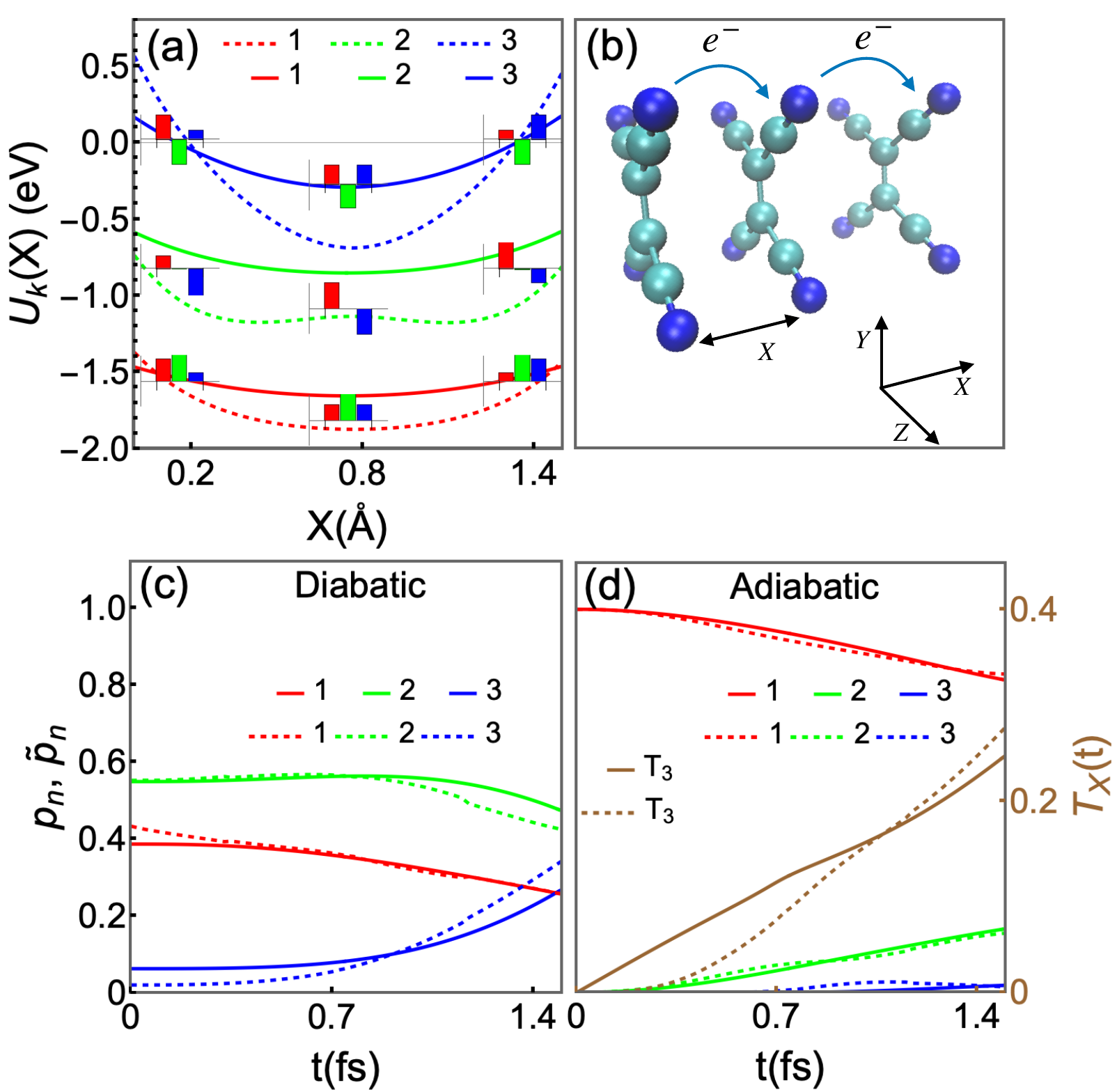}
\caption{(a) TCNE trimer, with one additional electron, and color coded localized states, with level diagram indicating $H_{nn}$. (b) The lowest three energies $U_k(z)$, from DFT (dashed) and effective state model (solid). The eigenstate amplitudes $d^{(k)}_n$ are superimposed, with colors matching molecule number in (a) and thin horizontal line $d=0$. (c) diabatic populations $p_n=|c_n|^2$ and (d) adiabatic populations $\tilde{p}_n=|\tilde{c}_n|^2$
in the same style for both methods. The adiabaticity measure $T_3$ is finally shown in (d), right axis.
\label{charge_transport}}
\end{figure}

\ssection{Charge transport in tdDFT}
%
To demonstrate the wide utility of the measure, we now move from effective state models to comprehensive molecular calculations using tdDFT \cite{Casida2006,AndreasCR2005,CurchodCPC2013}, 
widely used due to its good compromise between accuracy, computational cost and size scalability. While a wavefunction is initially not directly available in tdDFT, we will demonstrate how all required quantities for the adiabaticity measure can be evaluated from simulations, such that the measure can be leveraged for much of quantum chemistry and material science. For this, we consider a prototypical charge transport example, in an artificially moved tetracyanoethylene (TCNE) trimer \cite{DixonJACS1987,MCCORMAC20013287,LiaoJPCA2003,MilianCPC2005,ADAM2016311}, shown in \frefp{charge_transport}{a}, onto which we place an additional electron.

Within the \emph{ab initio} framework, setting up the measure requires the many-body states of the adiabats, their coefficients and the charge-localized diabats. The ground-state wave function is simply obtained as a Slater determinant of the molecular orbitals calculated from DFT, while the excited-states are constructed as linear combinations of excited singles using the Casida approach \cite{Casida2006, curchod2013trajectory}. For constructing diabats we employ constrained density functional theory \cite{QinJCP2006,VoorhisARPC2010,KadulCR2012}, limiting the additional electronic charge separately on individual TCNE molecules in our trimer. Both are implemented in NWChem~\cite{ValievCPC2010}. Finally, the time-dependent adiabatic coefficients $\tilde{c}_k(t)$ are obtained by solving the TDSE in the adiabatic representation $i\hbar \frac{\partial}{\partial t}  \tilde{c}_k(t)= U_k(t) \tilde{c}_k(t) - i\hbar\sum_m \kappa_{km}(t)\tilde{c}_m(t)
$, where $\kappa_{km}(t) = \langle\phi_k(t)|\frac{\partial}{\partial t}|\phi_m(t)\rangle$ are the non-adiabatic couplings. For this, we have used the trajectory surface hopping code SHARC \cite{Richter2011JCTC,Mai2018WCMS} with disabled hopping, interfaced with ORCA \cite{NeeseWcmsOrca2012,NeeseWcmsOrca2018}. More details on extraction of quantities from tdDFT can be found in the SI \cite{sup:info}.

Charge transport on the TCNE trimer can also be understood in an effective state model $\sub{\hat{H}}{eff}=\sum_{nm} H_{nm}(\mathbf{X}) \ket{n}\bra{m}$, where $\ket{n}$ imply the excess electron is localized on molecule $\#n$, $H_{nn}$ are effective on site energies, and $H_{nm}$ for $n\neq m$ are amplitudes for electron transfer. In general all matrix elements depend on the position $\mathbf{X}$ of the three monomers, as discussed in the SI \cite{sup:info}.
We fix the outer two TCNE molecules $R=7$~\AA\ apart, and initially locate the middle molecule $\#2$ closer to $\#1$, separated by $X(t=0)=2.9$~\AA, such that $V_{12}$ dominates. The amplitude $d^{(k=1)}_n$ to find the electron on molecule $n$ in state $k$ is then mostly shared between molecules $\#1$ and $\#2$ in the ground-state, populated initially. 

While moving molecule $\#2$ towards $X(\sub{t}{fin})=4.1$~\AA\ until $\sub{t}{fin}=1.5$ fs with constant exaggerated velocity, the excess electron in the ground-state now becomes localized on molecules $\#2$ and $\#3$.
Since the ground-state is mostly adiabatically followed, the charge on molecule $\#3$ increases, and the extent to which this has been aided by adiabaticity is captured by the measure $T_3$ shown in \frefp{charge_transport}{d}. 
Here $T_3(\sub{t}{f})/|\sub{c}{3}(\sub{t}{f})|^2=0.8$ only, at the final time $\sub{t}{f}$, due to the visible non-adiabatic effects. While this example was chosen for simplicity, the full power of the measure will unfold in ab-initio simulations of complex dynamics such as shown in \fref{fig_trimer_cradles}, columns 3,4. The proposed measure can then be leveraged in chemistry or material science to analyse \emph{ab initio} simulations based on MD and tdDFT. 
%

\ssection{Conclusions}
We quantified the extent to which adiabatic following of eigenstates is the root cause of quantum transitions in a non-eigenbasis. The correct functionality of our construction was  demonstrated using clear-cut cases where transitions are either fully due to adiabaticity or not at all. We then tackled examples from the regime of interest: complex quantum dynamics with multiple populated eigenstates and imperfectly adiabatic dynamics, in which a manual assessment of the relevance of adiabaticity becomes very challenging and an automatic one impossible. We finally introduced a pipeline for the evaluation of the measure in tdDFT. Quantification of the adiabatic contribution to a desired quantum process can now aid the design of nuclear dynamics for dye-sensitized solar cells \cite{zhang2017dye, ghosh1978merocyanine}, charge transfer to the acceptor \cite{Stier_Nonadiab_electransfer_JPCB} or the design of thin-film optical and optoelectronic devices \cite{malyshev2000intrinsic} but will also enable classification of complex quantum dynamics that is only partially adiabatic.

\acknowledgements
We thank the Max-Planck society for financial support under the MPG-IISER partner group program. The support and the resources provided by Centre for Development of Advanced Computing (C-DAC) and  the National Supercomputing Mission (NSM), Government of India are gratefully acknowledged.
RP is grateful to the Council of Scientific and Industrial Research (CSIR), India, for a Shyama Prasad Mukherjee (SPM) fellowship for pursuing the Ph.D (File No. SPM-07/1020(0304)/2019-EMR-I). The \emph{ab initio} calculations were carried out using the Paramshivay supercomputing facility at IIT BHU, India.

\section{Supplemental Information}

\section{Handling non-adiabatic transitions between eigenstates}

\begin{figure}[htb]
\includegraphics[width=0.99\columnwidth]{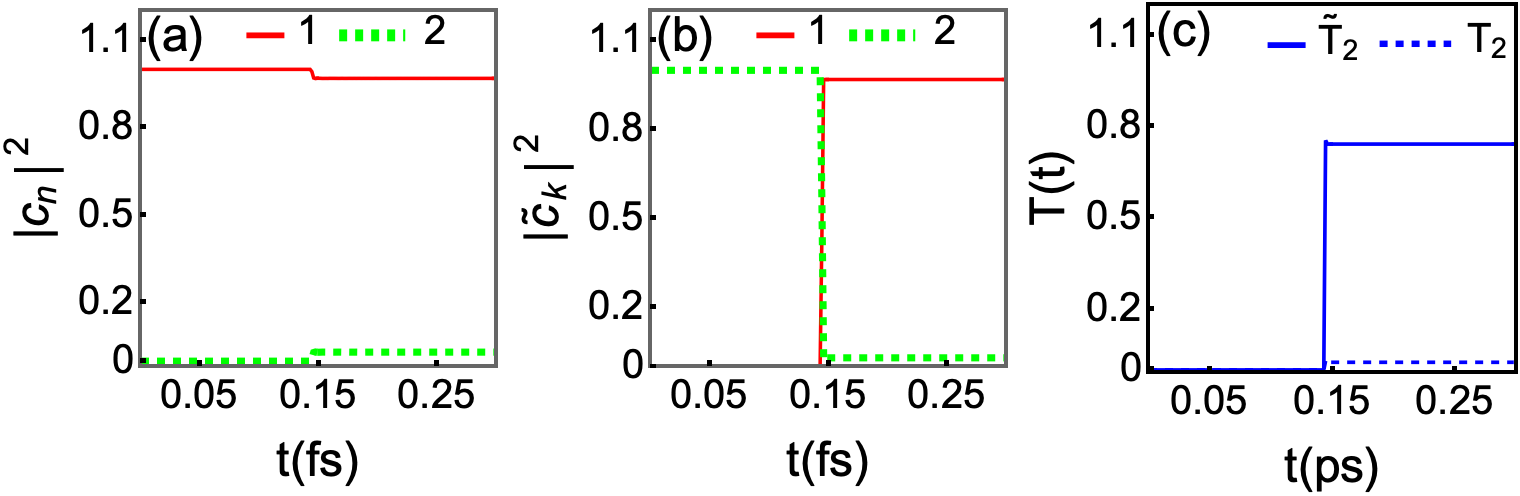}
\caption{\label{LZ_crossing} An example of sharp non-adiabatic Landau-Zener crossing in a two-level system to compare adiabaticity measures $T_2(t)$ and $\tilde{T}_{2}(t)$. (a) Diabatic populations. (b) Adiabatic populations. (c) Comparison of both measures for the dynamics described by Hamiltonian in \eref{LZHamiltonian}.}
\end{figure}

In the expression for the measure of adiabaticity \bref{pop_dot_no_phase}, the first term vanishes for completely adiabatic processes due to the derivative $\dot{a}_k(t)$. However, for dynamics that involve non-adiabatic changes, both terms become relevant for determining the adiabatic character of the transport.
To gain a better understanding of the significance of each term in \eref{pop_dot_no_phase}, we consider the example of a strongly non-adiabatic Landau-Zener crossing described by the Hamiltonian
\begin{align}\label{LZHamiltonian}
\hat{H} = \begin{pmatrix}
(t-\sub{t}{0})\frac{v}{2} & J\\
J & -(t-\sub{t}{0})\frac{v}{2} 
\end{pmatrix},
\end{align}
where $v$ is the rate of change of the diagonal elements in the Hamiltonian, $t$ is time and $J$ is the coupling between the two states. We compute the dynamics by solving the time dependent Schr\"odinger equation \eref{TDSE_diabatic_basis} using \bref{LZHamiltonian} and examine the resulting dynamics. We choose $v=1.8 \times 10^4$ cm$^{-1}$/ps with coupling strength is $J=4.4 \times 10^4$ cm$^{-1}$ and $t_0 = 0.14$ ps, which leads to a sharp non-adiabatic transition as shown in \frefp{LZ_crossing}{b}. In \frefp{LZ_crossing}{c}, we plot the measures $T_2(t)$ and $\tilde{T}_{2}(t)$ for the target state $X=2$, where the latter is calculated by dropping the first term in \eref{pop_dot_no_phase} to obtain
\begin{align}
\label{pop_dot_no_phase2}
\tilde{f}_n(t)= \sum_{k,k'}\tilde{c}^*_{k'}\: \tilde{c}_k \: \left( \dot{d}^{(k)}_n\: d^{(k')*}_n + d^{(k)}_n\: \dot{d}^{(k')*}_n \right) \bigg],
\end{align}
and from that $\tilde{T}_X(t)=  \int_0^t  dt' \tilde{f}_X (t')$. We observe a significant difference between the two constructions, with $\tilde{T}_{2}(t)$ falsely flagging an adiabatic transition into state 2 that has not happened.
This is remedied by the additional inclusion of terms in line 1 of Eq.~(4) of the main article.
%

\section{Alternative measures}

In many cases it will be of interest to boil down the set of functions $f_n$ defined in \eref{pop_dot_no_phase} into just a single number, to quantify the contribution of adiabaticity to a given state transition.
In the main article, we have concentrated on the temporal integral $T_X(t)=  \int_0^t  dt' f_X (t')$ of solely the function $f_X$ pertaining to a target site $X$.

Other variants will be possible and
the best choice may depend on the type of quantum dynamics for which one intends to characterise adiabaticity. Another option could be
\begin{align}
\bar{T}(t)&=\frac{1}{2}\sum_n  \int_0^t  dt' |f_n(t')|,
\label{totcrit_1}
\end{align}
which also takes into account potential intermediate transitions that the state has to evolve through before reaching $X$. This could be passage through the other monomers in the chain, for the examples in Fig.~2 of the main article. The expression \bref{totcrit_1} gives $K$ if a system makes a transition from one state $\ket{a}$ into a second state $\ket{b}$ through $K-2$ additional intermediate states entirely due to adiabatic following of a single eigenstate. It does treat transitions between all basis states $\ket{n}$ on equal footing and provides a time-averaged result for the entire duration $t$ of interest. 

However owing to the modulus, \bref{totcrit_1} also treats transitions \emph{into} some state equivalent to transitions \emph{out of} that state, which gives undesired results when some fast temporal changes in the Hamiltonian cause rapid oscillatory transitions back and forth between states, which average out in the dynamics but would add up in \eref{totcrit_1}, as demonstrated in \fref{fig_five_cradles}.

\section{Molecular aggregate model}
\begin{figure}[htb]
\includegraphics[width=0.99\columnwidth]{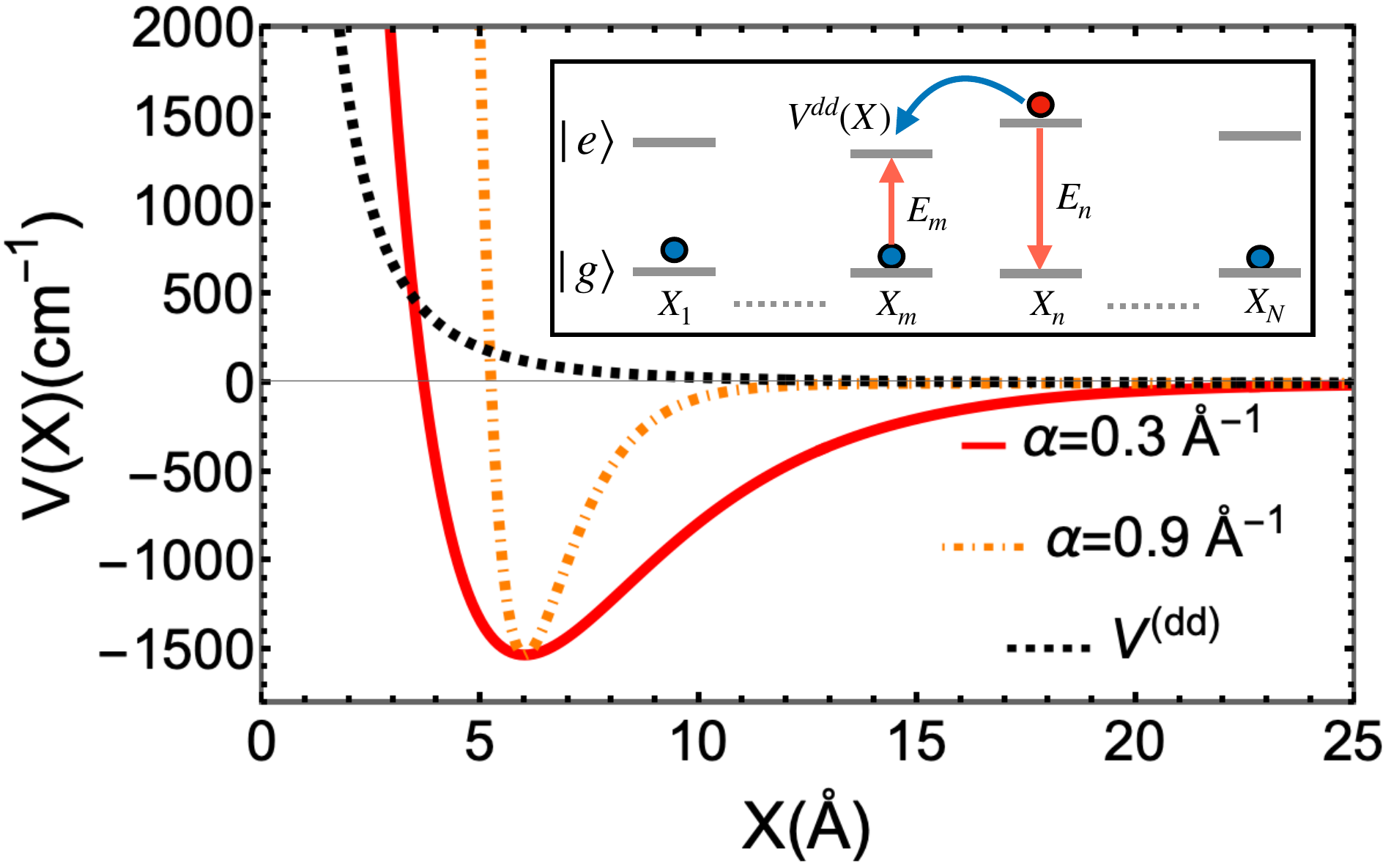}
\caption{ (a) Energy level schematic for a one dimensional chain of N molecules, with electronic ground state $\ket{g}$, excited state $\ket{e}$, dipole-dipole interaction $V_{dd}(X)$ and $E_n$ the site energy of the $n$'th molecule. (b) Inter-molecular Morse potential for $\alpha=0.3$ \AA$^{-1}$ (Red solid line) and $\alpha=0.9$ \AA$^{-1}$ (orange dot-dashed) and the strength of dipole-dipole interactions $V_{nm}^{(dd)}$ (black dashed).
\label{fig_molagg_geometry}}
\end{figure}

To model a molecular aggregate, we consider $N$ monomers of some molecular dye with mass $M$, arranged in a one dimensional (1D) chain along the $X$ direction. The positions of the molecules are given by $\mathbf{X}$ = ($X_1$, $X_2$ ,...., $X_N$) i.e., the $n$'th monomer is located at a definite, classical position $X_n$ and treated as a point particle. Adjacent monomers are assumed to interact with a Morse type potential
\begin{eqnarray}
\label{Morse_potential}
\sub V{mn}(\mathbf{X}) = D_e\Big[ e^{-2\alpha(X_{mn} - X_0)} - 2 e^{-\alpha(X_{mn} - X_0)} \Big],
\end{eqnarray}
where $D_e$ is the depth of the well, $X_{mn}=|X_n-X_m|$ the separation of monomers $n$ and $m$ with $X_0$ its equilibrium value and $\alpha$ controls the width of the binding potential. Examples are shown in \fref{fig_molagg_geometry}.

Additionally each monomer may be in an electronically excited state $\ket{e}$ or ground-state $\ket{g}$. Among the resultant many-body states, we restrict ourselves to the so-called single-exciton manifold, where just monomer number $n$ is excited, this state is denoted by $\ket{n}$. The excited state can then migrate to any other monomer via long-range dipole-dipole interactions. Altogether we thus have a classical Hamiltonian for molecular motion
\begin{eqnarray}\label{class_Hamiltonian}
\sub{H}{class} =  \sum_{n=1}^{N}  \frac{1}{2}M \dot{X}_n^2 + \sum_{n<m}\sub V{mn}(\mathbf{X}) ,
\end{eqnarray}
and a quantum mechanical one for excitation transport through dipole-dipole interactions
\begin{eqnarray}\label{Hamiltonian}
\hat{H}(\mathbf{X}) = \sum_{n=1}^{N}E_{n} \ket{n}\bra{n} + \sum_{\stackrel{n\neq m}{n,m}}\frac{ \mu^2}{X_{mn}^3} \ket{n}\bra{m},
\end{eqnarray}
where $E_n$ is the electronic transition energy of the $n$'th monomer and $\mu$ is the transition dipole moment. We find the system dynamics in a quantum-classical approach, where the motion of the molecules is treated classically using Newton's equations
\begin{eqnarray}
\label{newton_longit}
M \frac{\partial^2}{\partial t^2}{X}_n = - \frac{\partial}{\partial X_n}  U_k(\textbf{X}) - \sum_m \frac{\partial}{\partial X_n}V_{mn}(\mathbf{X}).
\end{eqnarray}
Here $U_k(\textbf{X})$ are the potential energy surfaces defined by using the adiabatic basis $\ket{\varphi_k[\textbf{X}(t)]}$, i.e. solving 
$H(\textbf{X}) \ket{\varphi_k[\textbf{X}(t)]}= U_k[\textbf{X}(t)] \ket{\varphi_k[\textbf{X}(t)]}  $. The dynamics of excitation transport is obtained by writing the electronic aggregate state as $\ket{\Psi(t)}=\sum_n c_n(t) \ket{n}$ and using Schr\"odinger's equation
\begin{eqnarray}
\label{TDSE_diabatic_basis}
i \hbar \frac{\partial}{\partial t} {c}_n = \sum_{m=1}^{N} H_{mn}[X_{mn}(t)] {c}_m.
\end{eqnarray}
Here $H_{mn}[X_{mn}(t)]$ is the matrix element $\bra{n}\hat{H}\ket{m}$ for the electronic coupling in \eref{Hamiltonian}. 

Projecting Schr{\"o}dinger's equation into the adiabatic basis provides further information about adiabatic evolution
\begin{eqnarray}
\label{TDSE_adiabatic_basis}
i\hbar \frac{\partial}{\partial t}  \tilde{c}_k(t)= U_k(t) \tilde{c}_k(t) - i\hbar\sum_m \kappa_{km} \tilde{c}_m(t).
\end{eqnarray}
Here $\kappa_{km}=\bra{\varphi_k(t)}\frac{\partial}{\partial t} \ket{\varphi_m(t)}$  are the non-adiabatic coupling vectors. We see that as long as the non-adiabatic coupling remain small, the system evolves adiabatically with eigenstate populations conserved $|\tilde{c}_k(t)|^2=|\tilde{c}_k(0)|^2$. Thus deviations of these populations from their initial value provide a measure of net non-adiabaticity while the size on non-adiabtic coupling terms provides a measure of instantaneous non-adiabaticity. In many simple cases these can be sufficient to assess to what extent a process is aided by adiabaticity, but not in the complex (and generic) ones that are our main focus.

 In the next section, we will show a wider collection of examples to illustrate also \eref{totcrit_1} and the difference to the construction of the main article, $T_X$.

\section{Thermal Motion of Molecules}
\label{thermal_motion_five}

\begin{figure}[htb]
\includegraphics[width=0.99\columnwidth]{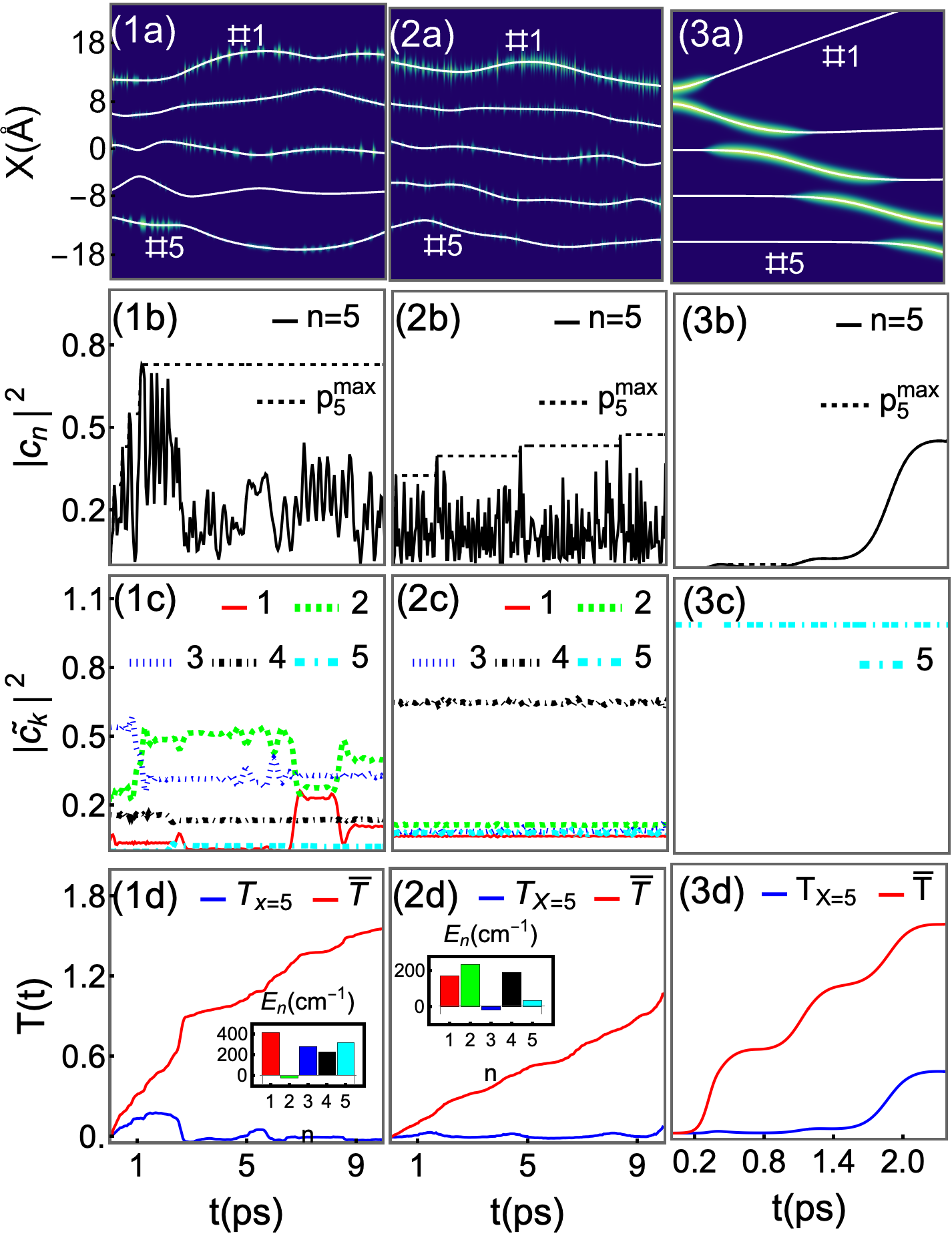}
\caption{ Exciton transport in molecular aggregates similar to \fref{fig_trimer_cradles}, with additional measure $\bar{T}(t)$ as described in \eref{totcrit_1}. The first column (1a-1d) is identical to the third column in \fref{fig_trimer_cradles} except the addition of $\bar{T}(t)$ here. The second column (2a-2d) shows the transport where the molecules are allowed to oscillate around their equilibrium position after been given random initial offset and velocity from a thermal distribution at room temperature. Finally, column (3a-3d) is for the complete adiabatic transport, similar to the second column in \fref{fig_trimer_cradles} of the paper, but for an aggregate of five molecules.
\label{fig_five_cradles}}
\end{figure}

To compare the measures $T_X (t)$ and $\bar{T}$ for the transport in thermally agitated molecular aggregates, we take an array of five monomers allowed to oscillate around their equilibrium separation after been given random initial offset and velocity from a thermal distribution at room temperature. The initial electronic state at $t=0$ is localized on the first site ($\#1$)
\begin{eqnarray}\label{single_site_inistate}
\ket{\psi(0)} =\ket{n=1}.
\end{eqnarray}
\fref{fig_five_cradles} shows the dynamics of excitation transport at temperature $T=300$ K using parameters listed in \tref{Table_Morse_parameters}.
\begin{table}[h]
\caption{Parameters for the Morse potential \eref{Morse_potential} and on-site disorder strength $\sigma_E$.}
\label{Table_Morse_parameters}
\begin{tabularx}{0.47\textwidth}
{  | >{\raggedright\arraybackslash}X | > {\centering\arraybackslash}X | > {\raggedleft\arraybackslash}X | }
 \hline
 First column & $D_e = 1528$ $cm^{-1}$ $\alpha = 0.5$ \AA$^{-1}$ & $\sigma_E= 300$ $cm^{-1}$  \\
 \hline
 Second column &  $D_e = 1528$ $cm^{-1}$ $\alpha = 0.5$ \AA$^{-1}$ & $\sigma_E= 150$ $cm^{-1}$  \\
 \hline
  Third column &  $D_e = 1528$ $cm^{-1}$ $\alpha = 0.5$ \AA$^{-1}$ & $\sigma_E= 450$ $cm^{-1}$  \\
\hline
\end{tabularx}
\end{table}

The first column in \fref{fig_five_cradles} is identical to the third column of \fref{fig_trimer_cradles} except the addition of $\bar{T}(t)$ for comparison with $T_X(t)$. Here wee see a slow steady increase in the measure $T_1(t)$, since both, population increase and decrease on the target site are cumulatively contributing. This problem is removed for measure $T_5(t)$, which is thus here more effective in identifying long-term useful adiabatic contributions to transport. 

For the second column (2a-2d) in \fref{fig_five_cradles}, the disorder is relatively small compared to the electronic dipole-dipole coupling. Due to the weak disorder, the excitation can reach the output-site with high amplitude at early times, before motion had a chance to impact dynamics.  In \rref{pant2020excitation} we quantify transport efficiency through the maximum of population on the output site (here 5) over the time of interest, shown as a black-dashed line in row (b). Probing the adiabaticity measures at the times where this maximum increases, gives a correctly constantly low contribution from adiabaticity only from measure $T_2$, not from $T_1$. The reason is as discussed for column one.

Finally for the last column (3a-3d), significant adiabatic transport can now be inferred directly from panels (3a), (3c), since exciton populations remain fairly adiabatic while almost the complete site population is transferred from \#1 to \#5. This leads to stepwise increases in the measure $\bar{T}$, not impacting $T_5$ before $t=1.7$ ps since the latter is based on site \#5 which was not yet involved.

We have seen that both measures give adequate results for certain regions in parameter-space, however care has to be taken with $\bar{T}(t)$ in \eref{totcrit_1} for cases where this adds up fast in- and out- transfer of population among basis states, that does however not yield a significant net transition when averaged over longer times. This is alleviated by measure $T_X(t)$ at the expense of being sensitive only to transitions into one specific state.

\section{tdDFT calculations}

\subsection{Methods}
\label{sec:technical}

Our \emph{ab-initio} calculations involve density functional theory (DFT), linear response time-dependent DFT (lr-tdDFT), constrained DFT (CDFT), and non-adiabatic molecular dynamics (NAMD), applied to a Tetracyanoethylene (TCNE) trimer, with monomer geometry taken from the manual of NWChem \cite{ValievCPC2010}. We use SHARC interfaced with ORCA for our NAMD to obtain the time-dependent adiabatic coefficients, $\Tilde{c}_k(t)$ and nuclear trajectories to utilize its unrestricted DFT approach, required due to an excess electron in TCNE trimer. Based on the obtained MD trajectory, we then recalculate adiabatic and diabatic states in NWChem using DFT for ground states and lr-tdDFT for excited states, to make use of existing utility codes for processing these.

As our objective here is proof of principle demonstration that all ingredients for the proposed measure can be extracted from first-principles simulations, we used NAMD merely to generate an artificial trajectory with the middle TCNE monomer moving at constant velocity without internal coordinate changes. For this, we started NAMD in the ground state and switched off surface hopping, and gave each atom in monomer \#2 the initial velocity $v_x = 0.8$~\AA/fs in the x-direction. This is unphysically high but chosen to enforce slight non-adiabatic transitions.
Subsequently, nuclei move due to the forces obtained from the instantaneous potential energy surfaces, which are calculated on the fly on a given nuclear time step by solving the Schr\"odinger equation. However internal deformation of the monomer was negligible, and during the time considered it just moved as a rigid object from the initial coordinate $X =  2.9$ {\AA} to the final $X = 4.1$ {\AA} during $t=1.5$ fs. Here the MD used a time step $dt = 0.025$ fs, while the co-propagated electronic dynamics needed for the instantaneous forces used  $dt'=0.001$ fs. As the system has an extra electron, all the states were taken to be doublets in the NAMD simulations. Here, the ground-state energy of the trimer geometry was taken as the reference energy, to avoid even shorter time steps for the electronic evolution. Spin-orbit couplings are not considered in the present calculations. 
 
 NAMD as described above provides us with nuclear trajectories $\mathbf{X}(t)$ and adiabatic coefficients, $\Tilde{c}_k(t)$. Based on these, we then used NWChem to regenerate the adiabatic states including the ground and excited states and diabatic states using CDFT, in order to make use of existing post-processing tools to calculate the determinant overlap between any two many-electron wavefunctions, see \eref{eq:phi_k_wfovrlp}. 
 
The 6-31G basis set is used for all of the above calculations, with exchange-correlation functional CAM-B3LYP \cite{YanaiCPL2004}, which combines the hybrid qualities of B3LYP and a long-range correction. CAM-B3LYP is known to yield atomization energies of similar quality to those from B3LYP, while also performing well for charge transfer excitations, which B3LYP underestimates enormously. 

\subsection{Diabats and orthogonalisation}
\label{subsec:diabats_ortho}

Finding the coefficients $d_n^{(k)}$ that are essential for the evaluation of the adiabaticity measure requires the many-body wavefunctions of the diabatic and adiabatic states, involving the ground and excited states. We obtain the diabatic states $|1\rangle, |2\rangle,$ and $|3\rangle$ by constraining the charge on the first, second, and third molecule using CDFT, respectively.
Note that each of these is still a many-body electron state involving all three molecules. While the constraint would work perfectly for a large separation of all molecules,
the algorithm works better for distances covering the distance range later required for dynamics, and diabatic states thus contain a correct admixture of slight delocalisation onto neighboring molecules.
We thus calculate diabats at $X\approx 3.5$~{\AA} for an equidistant trimer.

\begin{figure}[ht]
\includegraphics[width=0.9\columnwidth]{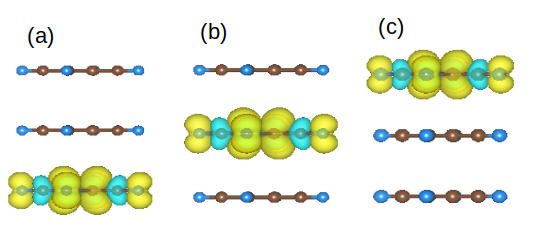}
\caption{(color online) Diabats: (a) $|1\rangle$, (b) $|2\rangle$, and (c) $|3\rangle$, corresponding to the states when the charge is localized on molecules 1, 2, and 3, respectively. Here, the spin densities are shown in terms of the iso-surfaces, yellow negative, cyan positive, together with a top view of the TCNE trimer shown in Fig.~3(b) of the main article.  
\label{fig:diabats}}
\end{figure}

However, at this distance, the diabats are not completely orthogonal, and were thus orthogonalised in a subsequent step:
Let $\{|\tilde {n}\rangle\}$ be the set of nonorthogonal but linearly independent basis functions, $\bm S$ their overlap matrix with the elements $S_{\tilde{n}\tilde{n}'} = \langle \tilde{n}|\tilde{n}'\rangle$, and $\{|n\rangle\}$ the orthonormalized functions we are looking for. We consider the case when the sets $\{|n\rangle\}$ and $\{|\Tilde{n}\rangle\}$ span the same subspace; then the orthogonal functions can be expanded in terms of the original ones as
\begin{equation}
    |n\rangle = \sum_{\tilde{n}=1}^N C_{\Tilde{n}n}|\tilde{n}\rangle,
    \label{eq:n}
\end{equation}
where $N$ is the number of diabats.

Using L\"owdin's symmetric orthogonalization scheme described in \rref{MayerIJQC2002}, states $\{\ket{n} \}$ are orthogonal for
\begin{equation}
    \bm C = \bm S^{-1/2}
    \label{eq:ortho_cond}
\end{equation}
where $\bm S$ is the overlap matrix of non-orthogonal diabats defined earlier.

 To verify whether the excess charge is fully localized on individual molecules within the CDFT, we have calculated the spin density for each diabat, shown in Fig.~\ref{fig:diabats} (a)-(c). It can be seen that for a given intermolecular distance of $X= 3.5$~\AA\ and for each diabat, the excess electronic charge, here showing up as spin density, is perfectly localized on individual TCNE molecules.

\subsection{Adiabatic construction}
\label{subsec:adiabats}
\begin{figure}[ht]
\includegraphics[width=0.7\columnwidth]{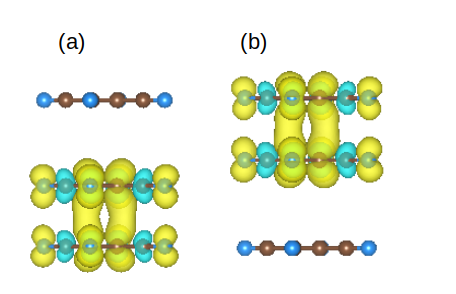}
\caption{(color online) The adiabatic ground state: (a) at the initial time $t=0$, while (b) shows it at the final time $\sub{t}{fin}$ as discussed in main text. Here, similar to \fref{fig:diabats}, the spin densities are shown in terms of the iso-surfaces, yellow negative, and cyan positive, together with a top view of the TCNE trimer.
\label{fig:adiabats}}
\end{figure}

The adiabatic states are obtained by DFT calculations on the TCNE trimer with an excess electronic charge without constraint. Initially, the second molecule was kept closer to the first molecule at a distance of 2.9~\AA, while keeping the third molecule at a distance of 7~\AA\ from the first. As shown in Fig.~\ref{fig:adiabats} (a) and (b) as an isosurface plot of the spin density, initially, the charge is localized on the first and second molecules, i.e. $|\psi(t=0)\rangle = \frac{1}{\sqrt{2}}(|1\rangle + |2\rangle)$ (Fig.~\ref{fig:adiabats} (a)), while at the final time $\sub{t}{fin}$, the motion of the middle molecule has changed the character of the ground-state such that the charge is now localized on the second and third molecules, i.e.~$|\psi(\sub{t}{fin})\rangle = \frac{1}{\sqrt{2}}(|2\rangle + |3\rangle)$ (Fig.~\ref{fig:adiabats} (b)).

\subsection{Projection of diabats onto adiabats}
\label{sec:wf_overlap}

The amplitude $d_n^{(k)} = \langle n|\phi_k(t)\rangle$ of a given diabat within the adiabatic state and its time evolution form an essential part of our proposed measure. We outline here briefly how it is obtained within the \emph{ab-initio} simulations. The actual antisymmetric many-electron wave functions $|\phi_k(t)\rangle$, also called the adiabatic states, and $|n\rangle$ are constructed as linear combinations of Slater determinants $\{|\Phi_l\}$ \cite{PhysRevLett.98.023001}
\begin{equation}
    |\phi_k(t)\rangle = \sum_{l=1}^{n_{CI}}f_{kl}(t)|\Phi_l(t)\rangle,
    \label{eq:phi_k_wfovrlp}
\end{equation}
where $f_{kl}$ are the configuration interaction (CI) coefficients, with index $k$ numbering the electronic states, and $n_{CI}$ is the number of elements in CI vectors.
Using these the amplitude becomes
\begin{equation}
    d_n^{k}(t) = \sum_l f_{kl}(t)\langle n|\Phi_l(t)\rangle.
    \label{eq:dnk_wfovrlp_2}
\end{equation}
For efficiently calculating the overlaps between any two many-electron wave functions, we employ the scheme described in
\rref{PlasserJCTC2016, werner2008nonadiabatic}.

\section{Simple model for charge transport}
%

We present a simple model to understand the transport of excess charge on the TCNE trimer in the main text. To use that, we have to first infer all its Hamiltonian matrix elements $H_{nm}$ from the more involved \emph{ab-initio} theory. 

\begin{figure}[htb]
\includegraphics[width=0.99\columnwidth]{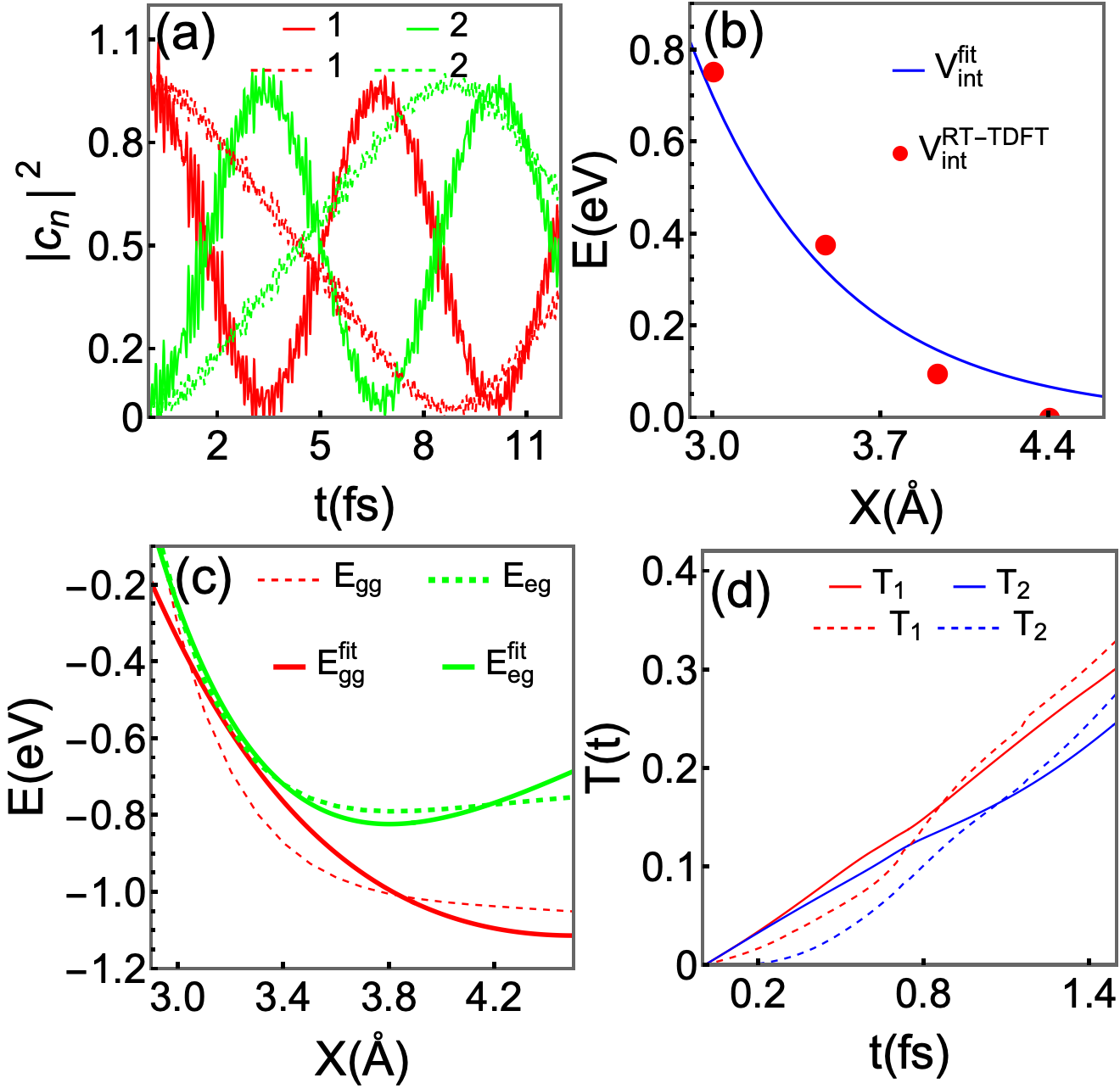}
\caption{ Excess electron transport in a TCNE dimer. (a) Time evolution of populations in a dimer at $3.5$ {\AA} (solid) and $4.0$ {\AA} (dashed). (b) 
Amplitude for electron transfer as a function of distance, obtained using RT-tdDFT (red dots) and fit with \eref{Vint_fit} (solid line). (c) Morse potential \eref{Morse_potential} (solid) fitted with DFT data (dashed). (d) Adiabaticity measures simple model (solid) and \emph{ab initio} calculations (dashed).
\label{Simple_model_comparision}}
\end{figure}

To infer off-diagonal coupling strengths $H_{nm}$ for $n\neq m$, which govern electron passage from monomer $n$ to monomer $m$, the time evolution of electron populations $p_n(t)$ in a TCNE dimer is calculated as a function of time for varying separation, using real-time dependent density functional theory (RT-tdDFT), as shown in \frefp{Simple_model_comparision}{a}. For this we start with the electron localized on just one monomer. The frequency of oscillations, obtained via a Fourier transform of populations $p_n(t)$, yields the coupling strength between the two localised electron states within the dimer at different intermolecular distances. These are then fit using an electronic interaction Hamiltonian,
\begin{eqnarray}\label{Vint_fit}
\sub{V}{int}(X) = Ae^{-BX},
\end{eqnarray}
as shown in \frefp{Simple_model_comparision}{b} with parameters in \tref{Table_fit_parameters}.

To infer the diagonal of the Hamiltonian in our effective model, we use DFT to compute the energy of the dimer for various distances with an extra electron constrained on a single molecule, $\sub{E}{eg}(X)$, and without an extra electron, $\sub{E}{gg}(X)$. The calculated energies are then fitted with a Morse potential, \eref{Morse_potential}, as shown in \frefp{Simple_model_comparision}{c}, with parameters in \tref{Table_fit_parameters}.

Finally, the on-site energies, $\sub{H}{nn}$, are given by, e.g.~
\begin{eqnarray}\label{On_site_energy}
\sub{H}{11} = \sub{E}{eg}(X_{12}) +  \sub{E}{eg}(X_{13}) +\sub{E}{gg}(X_{23}),
\end{eqnarray}
and similar for the other two energies. Off diagonal matrix elements are $\sub{H}{nm}=\sub{V}{int}(X_{nm})$.

\begin{table}[h]
\caption{Fit parameters for the Morse potential \eref{Morse_potential} and electronic interaction potential \eref{Vint_fit}. 
\label{Table_fit_parameters}}
\begin{tabularx}{0.47\textwidth}
{  | >{\raggedright\arraybackslash}X | > {\centering\arraybackslash}X | > {\raggedleft\arraybackslash}X | > {\raggedleft\arraybackslash}X | }
 \hline
$ \sub{D}{e}^{gg} = 0.91$ eV & $\alpha^{gg} = 0.40$ \AA$^{-1}$ & $\sub{X}{0}^{gg} = 4.5$ \AA & $A= 0.23$ eV \\
 \hline
$ \sub{D}{e}^{eg} = 0.82$ eV & $\alpha^{eg} = 0.75$ \AA$^{-1}$ & $\sub{X}{0}^{eg} = 3.8$ \AA  & $B=0.77$  \AA$^{-1}$ \\
\hline
\end{tabularx}
\end{table}
The results of our calculations are shown in \fref{charge_transport} of the main article, where we show the population dynamics and potential energy surfaces computed using both methods. The measure $T_X (t)$ and $T_1(t)$ are shown in \frefp{Simple_model_comparision}{d}, and we observe good agreement between both methods.
\providecommand{\noopsort}[1]{}\providecommand{\singleletter}[1]{#1}%

\end{document}